\documentstyle[times,epsfig,astrobib,useAMS]{mn2e}

\def\ni{\noindent}
\def\beb{\begin{thebibliography}{999}}
\def\eeb{\end{thebibliography}}
\def\bei{\begin{itemize}}
\def\eei{\end{itemize}}
\def\bef{\begin{figure}}
\def\eef{\end{figure}}
\def\ben{\begin{enumerate}}
\def\een{\end{enumerate}}
\def\beq{\begin{equation}}
\def\eeq{\end{equation}}
\def\ber{\begin{eqnarray}}
\def\eer{\end{eqnarray}}

\def\Bb{{\bf B}}
\def\pa{\partial}
\def\vb{{\bf v}}

\def\half{\frac{1}{2}}
\def\third{\frac{1}{3}}

\newcommand{\msun}{\mbox{{\rm M}$_{\odot}$}}
\newcommand{\mdot}{\mbox{$\dot{M}$}}

\title[neutron star field evolution]{Diamagnetic Screening of the
Magnetic Field in Accreting Neutron Stars II -- The effect of polar 
cap widening}
\author[Konar and Choudhuri]
{Sushan Konar$^{1}$\thanks{E-mail: sushan@cts.iitkgp.ernet.in; 
arnab@physics.iisc.ernet.in} and Arnab Rai Choudhuri$^{2}$\\
$^{1}$Indian Institute of Technology, Kharagpur 721302, India\\
$^{2}$Indian Institute of Science, Bangalore 560012, India}

\begin{document}

\date{}

\pagerange{\pageref{firstpage}--\pageref{lastpage}} \pubyear{2002}

\maketitle

\label{firstpage}

\begin{abstract}
Recently, we have proposed a model for the screening of the 
magnetic field of an accreting neutron star by the accreted 
material flowing from the polar regions towards the equator 
and sinking there underneath the surface \cite{ck02}. In this 
model it was assumed that the flow pattern remained stationary 
over time. However, as the surface magnetic field weakens, the 
accretion takes place over a wider region around the pole, 
making the flow more radial and isotropic. In the present work, 
we extend this two-dimensional model to include the time-dependence 
of the flow of the accreted material. The final radial flow is 
found to be less efficient in screening the magnetic field compared 
to the initial tangential flow. After an initial phase of rapid 
decay, the magnetic field slowly reaches an asymptotic value when 
the accretion becomes nearly isotropic and radial. Assuming the 
initial extent of the polar cap to be $\sim 5^0$--$10^0$, a simple 
geometric argument suggests that the magnetic field should decay 
by $3-4$ orders of magnitude before stabilizing to an asymptotic 
value, consistent with the magnetic fields observed in millisecond 
pulsars.
\end{abstract}

\begin{keywords}
magnetic fields--stars: neutron--pulsars: general--binaries: general
\end{keywords}

\section[]{Introduction {\label{sintro}}}

Observations suggest a connection between the low magnetic field 
of some neutron stars (binary/millisecond pulsars) and their being 
processed in binary systems, indicating an accretion-induced field 
decay in such cases. Radio pulsar data and their statistical analyses 
indicate that binary as well as millisecond and globular cluster 
pulsars, which almost always have a binary history and shorter spin 
periods, possess much lower field strengths, down to $\sim 10^8G$ 
compared to the isolated pulsars which appear to have much larger 
magnetic fields ($ \sim 10^{11}G$ -- $10^{13}G$) and longer spin 
periods (see Lorimer \citeyear{lorm01} and references therein). 
Evidently the accreting material induces a rapid dissipation of the 
magnetic field and also spins the neutron star up by bringing in 
additional angular momentum. 

We do not have any {\it a priori} knowledge of the interior currents
supporting the magnetic field of a neutron star. Depending on the 
generation mechanism, the field could either be supported by the crustal 
currents \cite{blnd79} or by the Abrikosov fluxoids of the proton 
superconductor in the stellar core. The evolution of the magnetic field 
in these two cases are expected to be different. Accordingly, two classes 
of models have been proposed for the evolution of the magnetic field in 
accreting neutron stars (see Bhattacharya \citeyear{db02}, 
Konar \& Bhattacharya \citeyear{kb01} for brief reviews) - one relating 
the field evolution to the spin evolution and the other attributing the 
field evolution to direct effects of mass accretion. Evidently, the 
starting point for these models is to assume different kinds of initial 
field configurations.

Another possible method of field reduction is to screen it by the 
accreting material. As the highly conducting accreting plasma settles 
and spreads onto the surface of the neutron star, it could produce a 
diamagnetic screening effect, burying the stellar field underneath
it. Interestingly, this mechanism would not depend on the location of 
the field in the stellar interior and would work irrespective of the 
nature of the interior currents. Though this mechanism was suggested 
quite early on \cite{bisn74,taam86,roma90,roma95}, it is only recently 
that the problem is being investigated in some quantitative detail. 

It turns out that the accretion flow in the polar region of a magnetized 
neutron star is extremely sensitive to various magneto-hydrodynamic
instabilities and it is difficult to make even a qualitative assessment 
of the effectiveness of screening without a full three-dimensional 
computation which is yet to be attempted. A one-dimensional plane-parallel 
model by Cummings et al. \citeyear{cumm01} has indicated that the 
diamagnetic screening is ineffective for field strengths above 
$\sim 10^{10}G$ as well as for accretion rates below $\sim 1\%$ of the 
local Eddington rate and at lower accretion rates the field can diffuse 
through the accreting matter. Recently, Melatos \& Phinney 
\citeyear{meph01} have calculated the hydromagnetic structure of a 
neutron star accreting symmetrically at both the magnetic poles as a 
function of the accreted mass, starting from a polytropic sphere plus 
a centrally located magnetic dipole and have evolved the configuration 
through a quasi-static sequence of two-dimensional Grad-Shafranov 
equilibria with increasing accreted mass. They find that the accreted 
material spreads equator-ward under its own weight, compressing the 
magnetic field into a thin boundary layer and burying it everywhere 
except in a narrow, equatorial belt. According to their calculation the 
magnetic dipole moment scales as $B_0^{1.3} \mdot^{0.18} M_a^{\-1.3}$ 
where $B_0$, $\mdot$ and $M_a$ are the initial field strength, the 
rate of accretion and the total accreted mass. 

In an earlier work (Choudhuri \& Konar \citeyear{ck02} - Paper~I 
henceforth) we have presented a two-dimensional model to demonstrate 
the mechanism of diamagnetic screening. A neutron star with strong 
magnetic field undergoes polar-cap accretion. The material falling 
onto the polar cap flows horizontally along the surface towards the 
equator. The flows coming from the two polar caps meet near the equator 
and then sink below the surface. A subsurface poleward counter-flow is 
expected in the layers immediately beneath the surface layer of equator-ward 
flow. Deeper inside, the material settles radially onto the core. We 
obtained an analytical expression describing this flow in Paper~I and 
studied the evolution of the magnetic field subject to such a flow, 
kinematically, using a numerical code. In the absence of magnetic buoyancy, 
we found that the magnetic field gets screened in the very short time 
scale of the surface flow. On the other hand, if magnetic buoyancy in 
the molten surface layers is included, the magnetic field is screened 
in the longer time scale of the interior flow. For a typical neutron 
star, these two time scales are estimated to be 1 year and $10^5$ years 
respectively. As magnetic buoyancy is likely to be important, we concluded 
that the magnetic field would be screened in the longer time scale of 
$10^5$ years. 

Observations of millisecond and binary pulsars also indicate that 
their magnetic fields undergo little or no field reduction after 
the phase of active mass accretion is over. It is believed that the 
decaying magnetic field reaches an asymptotic value and then stops 
decaying any further. One of the limitations of Paper~I has been 
that the calculations presented there threw no light on this aspect 
of the problem. We used a time-independent velocity field and 
consequently the magnetic field continued to decay indefinitely. In 
reality, the velocity field is expected to be highly time dependent. 
As the magnetic field becomes weaker and the magnetic pressure in the 
polar region drops, it is possible for the accreting material to flow 
through an increasingly larger region around the pole. In other words, 
the polar cap widens with time in an accreting neutron star. Eventually, 
the magnetic field becomes too small to be able to channelize the flow 
of the accreting material and the accretion becomes spherically symmetric. 
The aim of the present paper is to self-consistently incorporate this 
time dependence of the velocity field into the evolution of the magnetic 
field. We show that spherical accretion is less effective in screening a 
magnetic field compared to accretion through polar caps. Hence, as the 
polar cap widens and the accretion tends to be more isotropic, the 
screening of the magnetic field becomes less efficient and the 
decay of the magnetic field is arrested to a large extent. So, after
an initial rapid decrease, the magnetic field reaches a phase of very 
slow decay thereafter. This result gives us a clue as to why the 
magnetic field of the neutron star attains an asymptotic value and
does not decay any further. 

It should be mentioned here that similar conclusions were drawn
for the case of crustal currents undergoing accretion-induced ohmic 
dissipation \cite{kb97,kb99a} where a purely spherical accretion
was assumed to be operative at all times. In that case such behaviour
of the magnetic field depends crucially on the nature of the detailed 
structural physics of the neutron star crust. The same scenario prevails 
even for a spin-down induced expelled flux subject to ohmic dissipation 
in the crust \cite{kb99b}. These calculations augment the present work in 
which we trace the evolution of the magnetic field in the surface layers 
as the nature of the accretion changes with time. At later times, when 
the accretion becomes spherical the calculations presented in the earlier 
models would become relevant. 

The layout of this paper is as follows. The mathematical
formulation of the problem is presented in \S2. In \S3, we 
investigate the evolution of the magnetic field with time-independent 
flow, with different widths of the polar cap. In \S4, we obtain 
the time evolution of the surface field with a time-varying flow 
pattern, adjusting itself to the change in the field strength.
However, the results obtained in \S4 are based on an assumption of
the existence of a non-radial flow till very deep layers of the star,
which is rather unphysical and serves only the purpose of
demonstration. Therefore, in \S5 we go on to look at the problem where 
the material flow is confined to a thin crustal region. There has been
some disagreement regarding the flow pattern, first presented in
Paper~I, in particular about the presence of a reverse flow in the
deeper layers. Therefore, in \S6 we have demonstrated that the flow 
velocity proposed by us is quite general and incorporates flow
velocity without a reverse flow as well and have compared our results 
with those obtained by using this kind of flow pattern. Finally, 
we summarize our conclusions in \S7.

\section[]{Mathematical Formulation {\label{smath}}}

We follow the mathematical formulation developed in Paper~I and
present a brief summary here for the sake of completeness. The magnetic 
field of the neutron star evolves according to the induction equation:
\beq
\frac{\pa \Bb}{\pa t} 
= \nabla \times (\vb \times \Bb) 
  - \frac{c^2}{4\pi} \nabla \times (\frac{1}{\sigma} \nabla \times \Bb) \,,
\label{eq_ind}
\eeq
where $\sigma$ is the electrical conductivity of the medium (see, for 
example, Choudhuri \citeyear{chou}, Parker \citeyear{park}). Assuming an 
axisymmetric poloidal field, allowing us to represent the magnetic field 
in the form 
$\Bb = \nabla \times \left( A(r, \theta) \hat{{\bf e_\phi}}\right)$, 
we find that $A$ evolves according to the equation:
\beq
\frac{\pa A}{\pa t} + \frac{1}{s} (\vb. \nabla)(s A) 
= \eta \left( \nabla^2 - \frac{1}{s^2} \right)A \,, 
\label{eq_dadt}
\eeq
where $\eta = c^2/4\pi\sigma$ and $s = r \sin \theta$. Evidently, it
is the poloidal component of $\vb$ that affects the evolution of $A$. 
We integrate equation (\ref{eq_dadt}) subject to the following boundary 
conditions. The field lines from the two hemispheres should match 
smoothly at the equator, requiring $\partial A/\partial \theta = 0$ 
at $\theta = \pi/2$. To avoid a singularity at the pole, we should 
have $A = 0$  at $\theta = 0$. The magnetic field matches a potential
field at the surface of the neutron star, whereas the lower boundary
allows free advection of the field below. See Paper~I for details of
how these boundary condition are implemented.

Assuming a polar-cap accretion, we expect the material accumulated 
in the polar region to cause an equator-ward flow at the surface in 
both the hemispheres. We assume this flow to be confined in a shell 
which is primarily in the liquid part of the crust. Near the equator, 
the flows originating from the two polar regions meet, turn around and 
sink under the surface resulting in a pole-ward counter-flow immediately 
beneath the top layer. Eventually this material settles radially onto 
the core. It should be noted that the pole-ward counter-flow as well as 
the radially inward flow takes place mainly in the solid crystalline 
region. The new material enters the region of interest only at the polar 
cap and therefore the flow-velocity has non-zero divergence only near the 
polar region. Apart from that $\nabla.(\rho \vb)$ should vanish everywhere 
else.

\bef
\begin{minipage}{10.0cm}
\epsfig{file=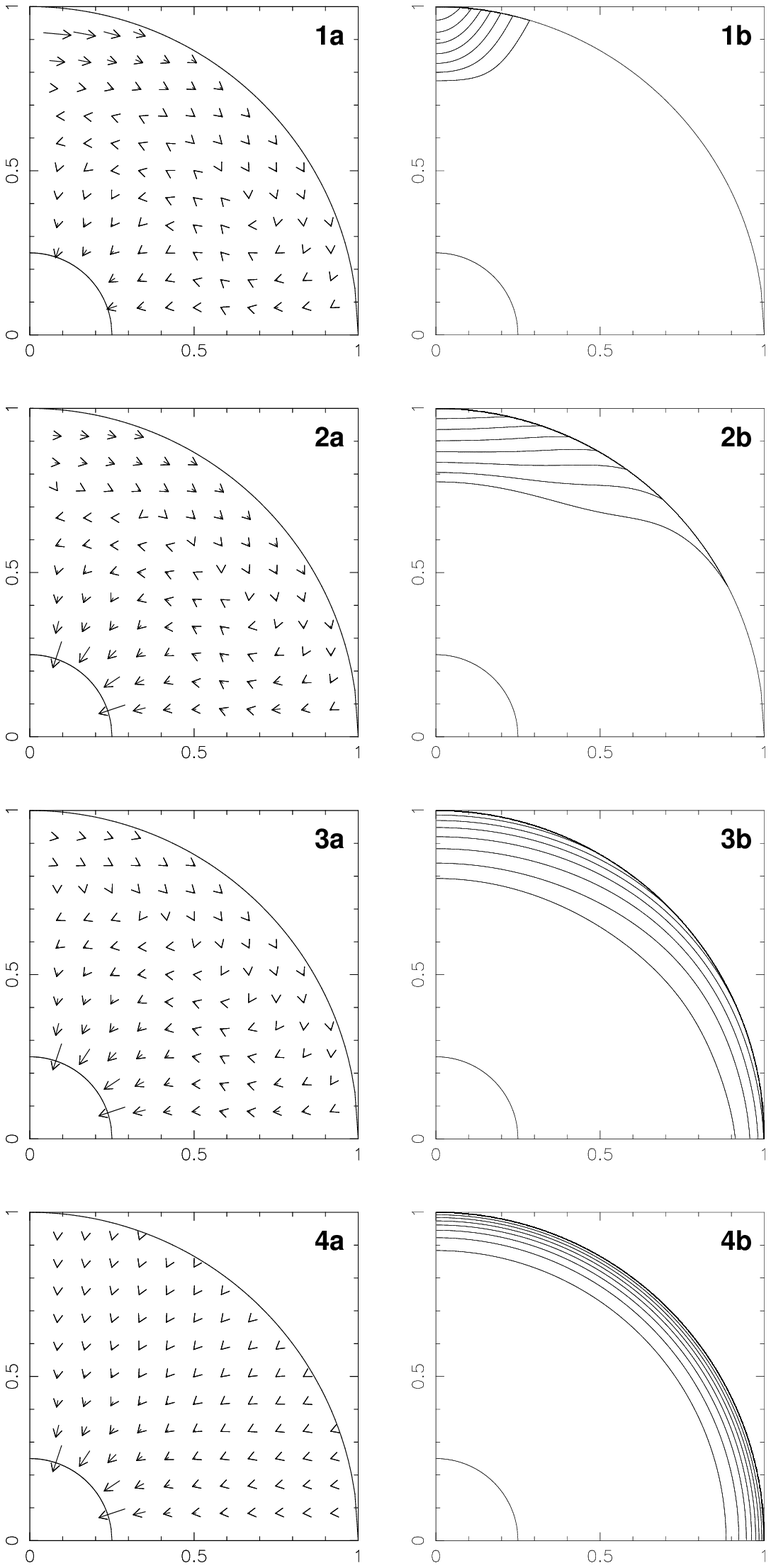,width=250pt}
\end{minipage}
\caption[]{Flow velocity, $\rho \vb$, and its divergence in a right-angular
slice ($0 \leq \theta \leq \pi/2$, $0.25 \leq r \leq 1.0$). We have
used $r_m = 0.75$, $r_b = 0.5$ for this picture. The panels marked
$1a, 2a, 3a, 4a$ correspond to flow velocities with $c = 0.0, 0.25,
0.5, 1.0$ and the panels marked $1b, 2b, 3b, 4b$ are the divergences
corresponding to those respectively.}
\label{fig01}
\eef

The analytic form of a velocity field, which has all the 
characteristics outlined above, is given below. 

\ni For $r_m < r < r_s$,
\ber
\rho v^{\bf 0}_r 
&=& K_1 \left(\frac{1}{3} r - \frac{1}{2} r_m 
    + \frac{\left( \frac{1}{2} r_m 
    - \frac{1}{3} r_s \right) r_s^2}{r^2} \right)
    e^{- \beta \cos^2 \theta} \,,\\
\rho v^{\bf 0}_{\theta} 
&=& \half \sqrt{\frac{\pi}{\beta}} K_1 \frac{(r - r_m)}{\sin \theta} 
    \mbox{erf} (\sqrt{\beta} \cos \theta) 
    \left(1 - e^{-\gamma \theta^2} \right) \,. 
\eer
\ni For $r_b < r <r_m$,
\ber
\rho v^{\bf 0}_r 
&=& K_2 \, e^{-\beta \cos^2 \theta} \nonumber \\
&\times& \left(\third r - \half (r_m + r_b) + \frac{r_m r_b}{r} 
         + \frac{\left(\frac{1}{6} r_b 
         - \half r_m \right) r_b^2}{r^2} \right) \nonumber \\
&-& \left(\third r - \half (r_m + r_b) + \frac{r_m r_b}{r} 
    + \frac{\left( \frac{1}{6} r_m 
    - \half r_b \right) r_m^2}{r^2} \right) \nonumber \\
&\times& \half \sqrt{\frac{\pi}{\beta}} K_2 \,,\\
\rho v^{\bf 0}_{\theta} 
&=& \frac{K_2}{2 \sin \theta} \sqrt{\frac{\pi}{\beta}} 
    \left( r + \frac{r_m r_b}{r} - r_b - r_m \right) \nonumber \\
&& \times \left(\mbox{erf}(\sqrt{\beta} \cos \theta) - \cos \theta \right) \,.
\eer
\ni For $r<r_b$,
\beq
\rho v^{\bf 0}_r = - \frac{K_3}{r^2} \,,
\label{eq_vrint}
\eeq
\beq
\rho v^{\bf 0}_{\theta} = 0 \,.
\eeq
Notice that the parameter $\gamma$ in 
$\vb^{\bf 0} (\beta, \gamma, r_b, r_m, r_s)$ defines the size of the 
polar cap, the angular extent being given by $\sim \gamma^{-1/2}$. 
Whereas, the parameter $\beta$ determines the angular width of the 
down-flow that sinks inward below the equator. The flow is 
predominantly horizontal in the surface layer $r_m < r <r_s$. 
In the layer  $r_b < r <r_m$ immediately below that, there is 
a pole-ward counter-flow. In this layer, the flow also tends to 
become more and more radial as it travels toward the deeper layers. 
Finally, below $r_b$, the material velocity becomes completely radial, 
flowing inward.

The flow in the innermost layers is nothing but the radial
compression experienced by the deeper layers of the star due to
an increase in the mass. Therefore, the magnitude of the velocity
is related to the rate of accretion by the condition
$K_3 = \mdot/{4 \pi}$. Ensuring continuity across $r = r_m$ and
$r = r_b$ we can relate $K_1$ and $K_2$ to $K_3$ and find that
$\nabla. (\rho \vb^{\bf 0}) = 0$ everywhere below $r = r_m$. However,
the divergence in the uppermost layer above $r = r_m$ is :
\ber
\nabla. (\rho \vb^{\bf 0})
&=& K_1 \, e^{-\gamma \theta^2 - \beta \cos^2\theta} \,
   \frac{r - r_m}{r} \nonumber \\
&& \times \left(1 + \sqrt{\frac{\pi}{\beta}} \frac{\gamma\theta}{\sin \theta}
   e^{\beta \cos^2\theta} \mbox{Erf}(\sqrt{\beta} \cos\theta) \right)
\eer
providing for a source of material only around the polar region in 
the upper layer. The panels $1a, 1b$ of Fig.~\ref{fig01} corresponds
to this flow velocity and its divergence, defined above. (In Fig.~1 
of Paper~I we have shown this velocity field 
$\vb^{\bf 0} (\beta= 10.0, \gamma=10.0, r_m =0.75, r_b = 0.5)$
and its divergence $\nabla .(\rho \vb^{\bf 0})$. Please note that there 
is a typographical error in Paper~I, where the values of both $\beta$ 
and $\gamma$ are mistakenly given as 1.0 instead of 10.0.). 

As the magnetic field in the polar region weakens due to screening, 
(a) the polar cap widens and (b) the velocity field tends to become 
more radial. Eventually, when the magnetic field is sufficiently weak, 
the accretion becomes isotropic and spherical instead of being polar, 
with the flow velocity becoming completely radial everywhere. We 
represent both of these effects through a single parameter $c$ taking 
values within the range $(0,1)$. A larger value of $c$ implies a wider 
extent of the polar cap and a more radial inflow. The effect of the 
widening of the polar cap can be made self-consistent by relating 
$\gamma$ and $c$ in the following manner:
\beq
\gamma = (\theta_{\rm min} + c \Delta \theta)^{-2}.
\label{eq_c}
\eeq
Since $\gamma^{-1/2}$ gives the angular width of the polar cap, we 
can see that $c=0$ corresponds to an angular width of 
$\sim \theta_{\rm min}$, whereas $c=1$ corresponds to 
$\theta_{\rm min} + \Delta \theta$. In this work, we have assumed 
$\theta_{\rm min} = 0.2$ and $\Delta \theta = 1.4$ such that the 
polar cap angle has a range of $\sim 10^o-90^o$. Through $\gamma$, 
the flow velocity $\vb_0 (\beta, \gamma, r_b, r_m, r_s)$ too depends 
implicitly on $c$. In order to make the flow velocity more isotropic 
with the widening of the polar cap, an isotropic part is added to 
the velocity field. Therefore, the expression for the velocity 
field, for a time-dependent magnetic field strength and hence a 
time-dependent polar cap area, is given by (only showing the
dependence on $c$)
\beq
\vb = (1-c) \, \vb^{\bf 0}\left(\gamma(c)\right) + c \vb^{\bf 1} 
\label{eq_vmod} \,.
\eeq
Here $\vb^{\bf 1}$ is the purely isotropic part given by 
\beq
\rho \vb^{\bf 1} 
= -\frac{K_3}{r^2} 
  \left( 1 - \exp \left(5 \frac{r_s - r}{r_s - r_m} \right)
  \right)\,. 
\label{eq_v1}
\eeq
It is evident that $\rho \vb^{\bf 1}$ vanishes at the surface and 
reaches an asymptotic value of $-K_3/r^2$ somewhere beneath the 
surface. It should be noted here that we are concerned with 
velocities inside the neutron star and not with those with which 
the accreting material may hit the surface of the star (which is 
certainly not zero at the surface). As freshly-accreted material
makes the new surface up, the original surface moves deeper down. 
The above expressions define the material flow with respect to the 
surface at any given instant of time. Note that equations \ref{eq_vmod}
and \ref{eq_v1} imply a radial velocity of $-K_3/r^2$ below $r=r_b$, 
which is independent of $c$. Thus the rate of inflow of material to 
the neutron star core (which is determined by $\dot{M}$ and is related 
to $K_3$) does not change with $c$. 

Fig.~\ref{fig01} shows the velocity field and its divergence for 
different values of $c$. It can be seen that the polar cap (indicated 
by the region where $\nabla.(\rho \vb)$ is substantially different 
from zero, implying deposition of accreted material) becomes larger 
with increasing values of $c$ and the flow velocity also becomes more 
isotropic. Finally, for $c = 1.0$, the velocity field becomes fully 
isotropic and radially inward everywhere. 

We now discuss how $c$ at a certain instant of time depends on 
the magnetic configuration at that instant. The structure of the 
magnetosphere around an accreting neutron star is a well-studied 
subject (see, for example, Shapiro \& Teukolsky \citeyear{shap}). 
The extent of the polar cap area is limited by open field lines on 
the surface of the star. The last open field line is the one that goes 
through the Alfv\'{e}n radius, $r_{\rm A}$, in the equatorial plane. 
Therefore, using the equation for dipolar field lines, we have
\beq
\sin \theta_{\rm P} = \left(\frac{r_s}{r_{\rm A}} \right)^{1/2}
\eeq
where $\theta_{\rm P}$ is the angle between the last open field line 
and the magnetic axis, and therefore, the angular extent of the polar 
cap. At the Alfv\'{e}n radius of an accreting system, the ram pressure 
of the in-falling material equals the pressure of the magnetic field. 
Now, the ram pressure at the Alfv\'{e}n radius is
\beq
P_{\rm ram} = \frac{1}{2} \rho(r_{\rm A}) V^2(r_{\rm A}) \,,
\eeq
where $\rho(r_{\rm A})$ is the density and $V(r_{\rm A})$ is the
velocity of the accreting material at Alfv\'{e}n radius. Assuming 
matter to be in approximately free fall, we have 
$V(r_{\rm A}) = (2GM/r_{\rm A})^{1/2}$. The density of accreting 
material at Alfv\'{e}n radius is 
$\rho(r_{\rm A}) = \mdot/4 \pi V(r_{\rm A}) r_{\rm A}^2$. Since the 
field is dipolar in nature, the field strength at the Alfv\'{e}n
radius is given by
\beq
B(r_{\rm A}) = \left(\frac{r_s}{r_{\rm A}} \right)^3 B_s \,,
\eeq
where $B_s$ is the field strength at neutron star surface. Equating 
the ram pressure and the magnetic pressure, one obtains
\ber
r_{\rm A} &=& (2GM)^{-1/7} r_s^{12/7} B_s^{4/7} {\mdot}^{-2/7}\,.
\eer
Assuming the rate of accretion (\mdot) and other stellar parameters 
($r_s, M$) to be constant, we then have the following relation between 
the angular extent of the polar cap and the surface field strength:
\beq
\sin \theta(t)_{\rm P} \propto B_s(t)^{-2/7} \,. 
\label{eq_tht1}
\eeq

In our model, the velocity field automatically adjusts to the
change in the extent of the polar cap through a variation in
$\gamma$. It follows from equation (\ref{eq_c}) that the opening angle of 
the polar cap $\theta(t)_{\rm P}$, given by $\gamma^{-1/2}$, is
\beq
\theta(t)_{\rm P} = \theta_{\rm min} + c(t) \, \Delta \theta \,.
\label{eq_tht2}
\eeq
Assuming $\theta_{\rm min}$ to correspond to the initial surface value 
of the magnetic field $B_s(t=0)$, it follows from equation
(\ref{eq_tht1}) and 
equation (\ref{eq_tht2}) that
\beq
\frac{\sin (\theta_{\rm min} + c(t) \, \Delta \theta) }{\sin \theta_{\rm min}} 
= \left[ \frac{B_s(t)}{B_s(t=0)} \right]^{-2/7}\,. 
\label{eq_ct}
\eeq
It may be noted that equation (\ref{eq_ct}) is valid only during the 
phase of accretion when the opening angle increases with the weakening 
magnetic field. When $c(t)$ equals one the accretion is completely
spherical implying the extent of the polar cap to be $\pi/2$. At this
point, the value of $c(t)$ is frozen and no further dependence of the
magnetic field is included in $c(t)$ and the flow velocity profile.

To obtain the evolution of the magnetic field with time, we solve 
equation (\ref{eq_dadt}) with the velocity field described above. 
This velocity field, in turn, depends on the strength of the magnetic 
field through the parameter $c$. In most of our calculations, we have 
included magnetic buoyancy in the manner prescribed in Paper~I, i.e, 
by including a radially upward velocity in the top layer. It has 
already been mentioned in Paper~I that the numerical program used for 
this work has been adopted from that developed in the context of solar 
MHD which formed the basis of several calculations involving solar 
magnetic fields \cite{dikp94,dikp95,chou95,chou99,nand01,nand02}. The 
details of the numerical scheme can be found in the Appendix of Paper~I.

\bef
\begin{center}{\mbox{\epsfig{file=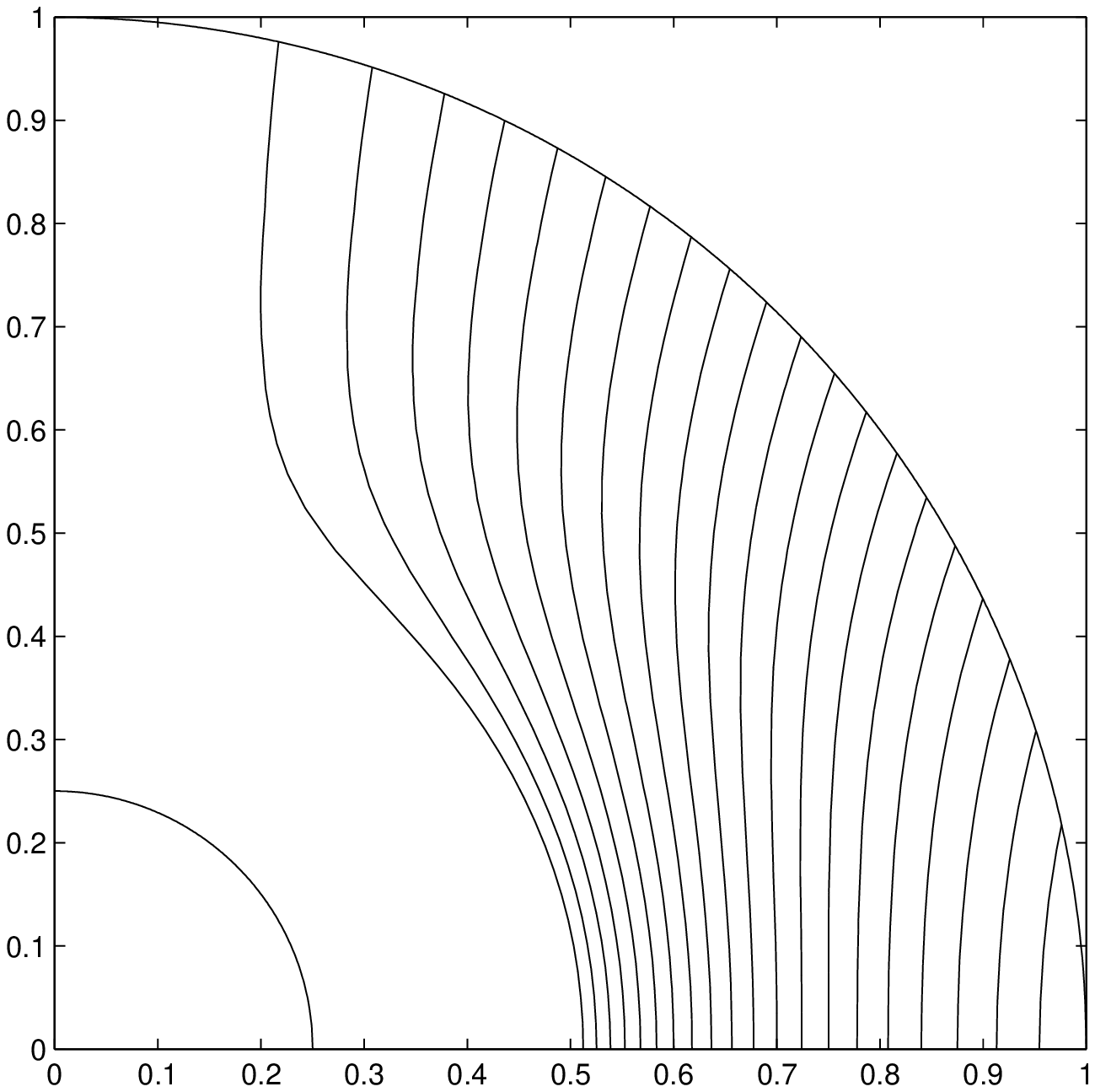,width=200pt}}}\end{center}
\caption[]{Initial field configuration, assuming a dipole confined to
an outer region defined by $0.25 \leq r \leq 1$ and 
$0 \leq \theta \leq \pi/2$.}
\label{fig02}
\eef

\bef
\begin{center}{\mbox{\epsfig{file=fig03.ps,width=150pt,angle=-90}}}\end{center}
\caption[] {Evolution of the mid-latitude surface field with time. The curves 
$a, b, c, d, e$ correspond to $c = 0.0, 0.25, 0.5, 0.75, 1.0$ respectively.
$\eta = 0.05$ and $v_{\rm mb} = 50$ for all the curves.}
\label{fig03}
\eef

\section{Modified Flow Velocity {\label{smodif}}}

\bef
\begin{center}{\mbox{\epsfig{file=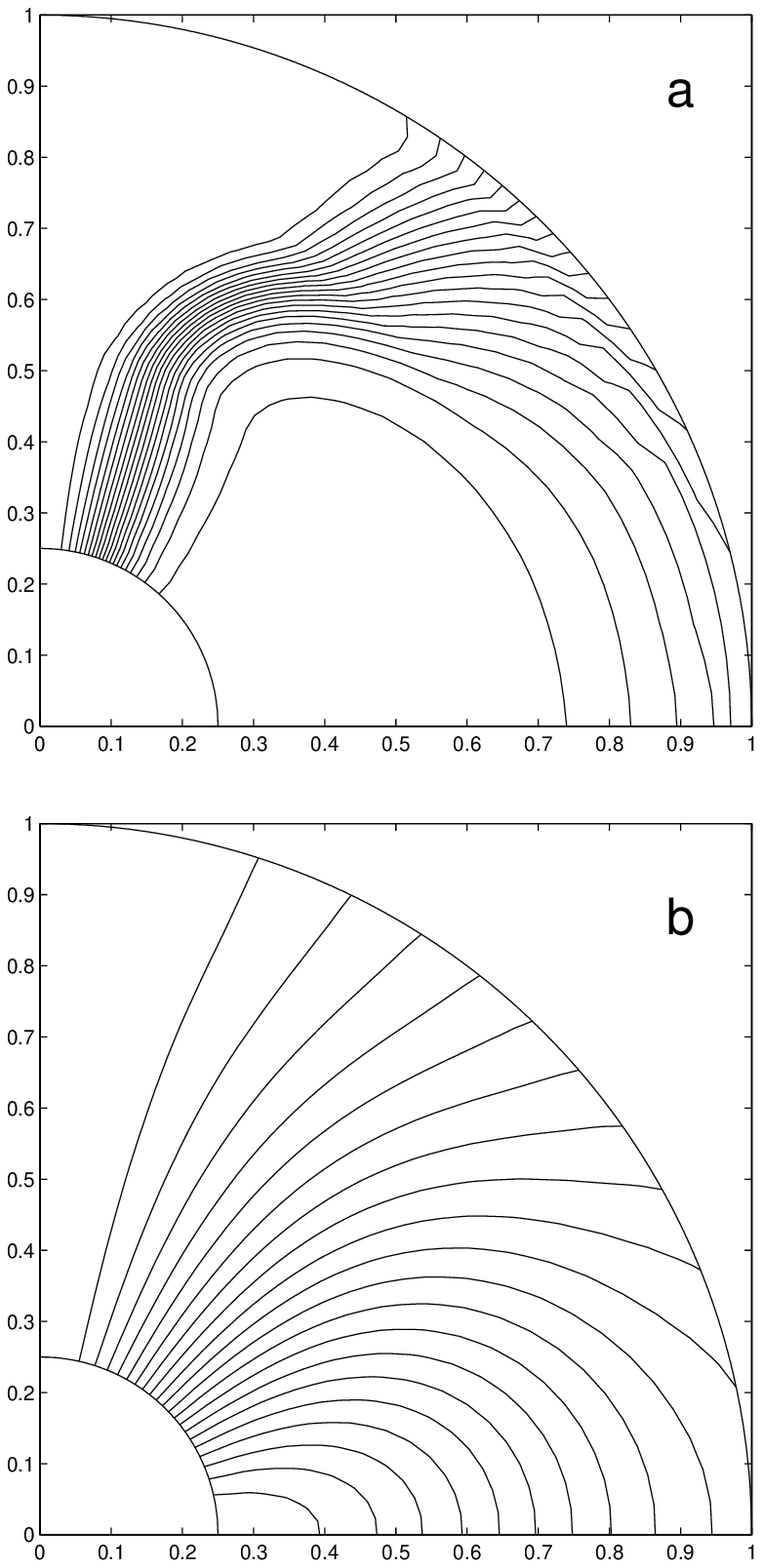,width=200pt}}}\end{center}
\caption[]{Field configuration at $t = 1.0$ starting from the initial 
configuration shown in Fig.~\ref{fig02}. The panels marked $a,b$ 
correspond to $c = 0.0, 1.0$ respectively (corresponding to the curves
$a$ and $e$ of Fig.~\ref{fig03}). In both the cases, $\eta=0.05$ and 
$v_{\rm mb} = 50$ are assumed.}
\label{fig04}
\eef

Before investigating the evolution of the magnetic field for a 
time-dependent flow velocity (resulting from a time-dependent size 
of the polar cap), we look at some representative cases with steady 
flow patterns corresponding to different sizes of the polar cap. For 
this, we use the modified form of the material flow velocity given 
by equation (\ref{eq_vmod}). The results of this section give an 
indication as to how the variation in the flow pattern affects the 
evolution of the magnetic field. 

As in Paper~I, we take the radius of the neutron star to be the 
unit of length. The unit of time is then fixed by equating the constant 
$K_3$ in the expression of velocity to unity. Our previous estimate
shows that the equator ward flow at the surface takes place only through 
a narrow layer of 100 m, which is $\sim 1\%$ of the stellar radius
(Paper~I). It is difficult to handle such a narrow layer in a numerical 
scheme. Moreover, it has been seen that most of the qualitative behaviour 
can be understood quite well by using a wider flow layer. Consequently, 
we have assumed $r_m = 0.75$, $r_b = 0.5$ for our calculations in this 
section and the next. Representative results with a narrow layer will be 
presented in \S5. We take the diffusivity to be $\eta = 0.05$. As shown 
in Paper~I, the magnetic field is more or less frozen in the fluid for 
this value of $\eta$ and the results do not change on decreasing $\eta$ 
any further. The parameter $v_{\rm mb}$ prescribing the magnetic buoyancy 
is set to the value 50.0. For all our calculations we assume $\beta$ to
be equal to $10.0$.

\ni We assume the initial currents to be entirely confined to the
outer surface layers of the star and start with an initial magnetic 
field configuration shown in Fig.~\ref{fig02}. This configuration
is allowed to evolve in time with steady flow velocities corresponding to 
different sizes of the polar cap. The flow patterns for different extents
of polar cap are obtained by taking different values of the parameter $c$
(as depicted in the velocity fields shown in Fig.~\ref{fig01}).
Fig.~\ref{fig03} shows how the magnetic field at a point on the 
surface at the mid-latitude decays with time. It is clear that
the decay slows down with increasing $c$, i.e.\ with increasing 
width of the polar cap. As the velocity field becomes more radial, 
it is less effective in screening the magnetic field. Fig.~\ref{fig04} 
shows the magnetic field configurations at time $t=1.0$ for $c=0$ and
$c=1.0$. It is evident that when $c$ is zero, the effect of a reverse
flow in the interior layers is quite prominent. Whereas, when $c$
equals unity, i.e.\ the accretion is purely spherical, the field
configuration shows only the effect of a radial compression. The fact 
that a more radial velocity field is less effective in screening the 
magnetic field, helps us to understand the results obtained with 
time-dependent velocity field more readily.

\bef
\begin{center}{\mbox{\epsfig{file=fig05.ps,width=150pt,angle=-90}}}\end{center}
\caption[]{Evolution of the mid-latitude surface field with time for
a time-dependent flow pattern (solid curve) with $\eta = 0.05$ and 
$v_{\rm mb} = 0$. The dashed curve shows the evolution of $c(t)$ with
time.}
\label{fig05}
\eef

\bef
\begin{center}{\mbox{\epsfig{file=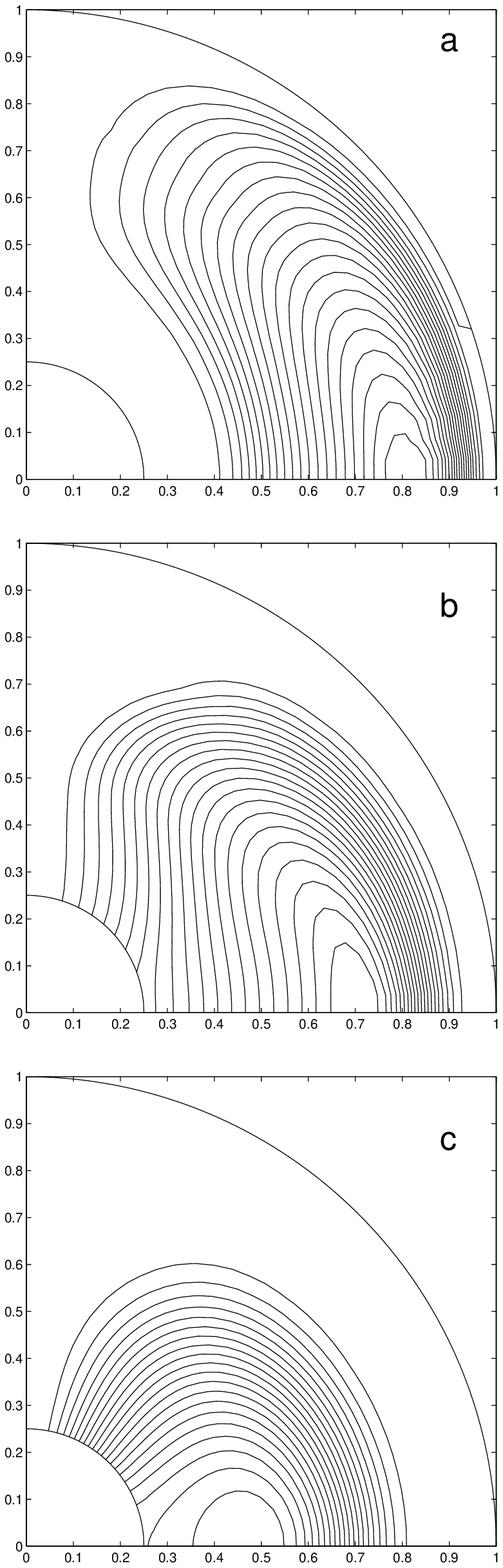,width=200pt}}}\end{center}
\caption[]{Field configuration at intermediate times starting from 
the initial configuration shown in Fig.~\ref{fig02}. The panels marked
$a,b,c$ correspond to $t = 0.015,0.05,0.1$ respectively.}
\label{fig06}
\eef

\section{Time-Dependent Flow Velocity {\label{stime}}}

\bef
\begin{center}{\mbox{\epsfig{file=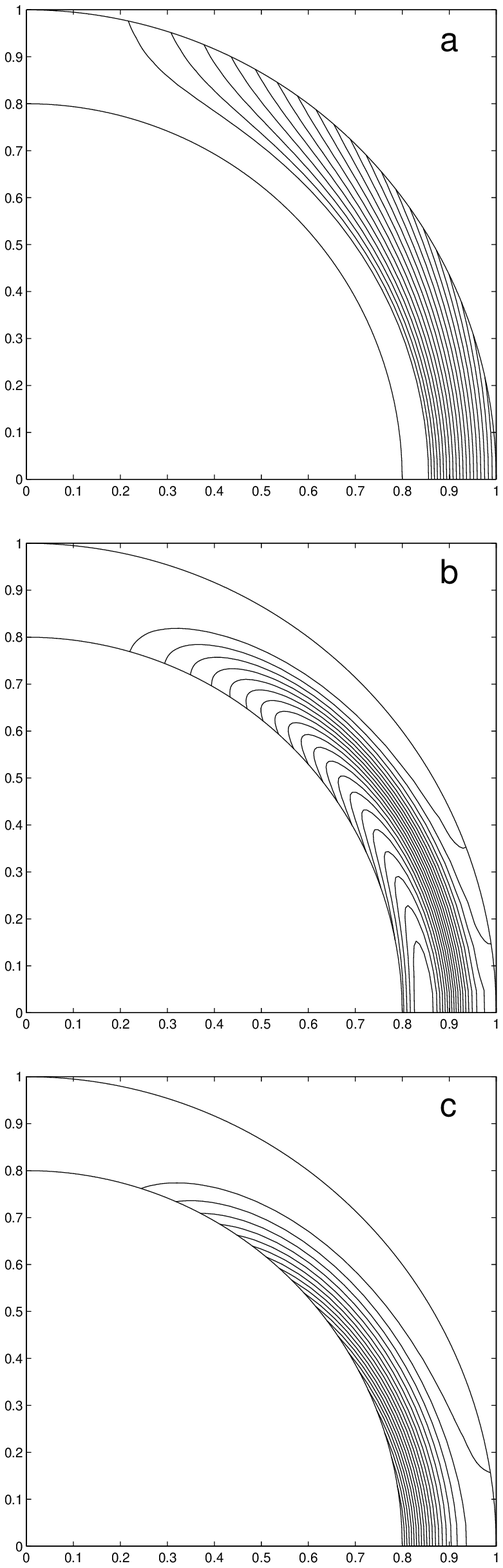,width=200pt}}}\end{center}
\caption[]{Field configuration at intermediate times starting from
panel $(a)$, confined to a region with $r \leq 0.8$ for $\eta = 0.05$ 
and $v_{\rm mb} = 50$. The panels marked $b,c$ correspond to 
$t = 0.05,0.1$ respectively.}
\label{fig07}
\eef

\bef
\begin{center}{\mbox{\epsfig{file=fig08.ps,width=150pt,angle=-90}}}\end{center}
\caption[] {Evolution of the mid-latitude surface field with time for
a time-independent flow pattern (solid curve). The initial field 
configuration is confined to a thin region as shown in the panel (a) 
of Fig.~\ref{fig07}. $\eta = 0.05$ and $v_{\rm mb} = 50$ has been 
assumed here. The dashed curve shows the evolution of $c(t)$ with
time.}
\label{fig08}
\eef

As accretion progresses, the magnetic field becomes weaker (due 
to screening, and also due to the ohmic dissipation in case of 
the crustal currents) and consequently the polar cap widens 
according to equation (\ref{eq_tht1}). This translates to equation 
(\ref{eq_ct}) which gives the time-dependence of the parameter $c$, 
in terms of the time-dependence of the field strength at the surface. 

Fig.~\ref{fig05} shows the evolution of the magnetic field at a point 
on the surface at the mid-latitude with time. It is clear that the
decay slows down with time. As the polar cap widens out with time,
the parameter $c$ also increases. Since the magnetic field decay is 
slower for larger $c$, as we saw in the previous section, the evolution 
seen in Fig.~\ref{fig05} is quite expected. Fig.~\ref{fig06} shows 
the magnetic configurations at three instants of time, demonstrating 
the physical nature of the distortion and effective screening of the 
magnetic field. As expected from the results of the previous section, 
we see that gradually the field configuration shows the effect of a 
reverse interior flow less and less. 

One of the important characteristics of binary and millisecond
pulsars is that their surface field strength is weaker by a factor 
of about $10^3$ -- $10^4$ compared to that of the normal pulsars.
Therefore, it has been stressed that the magnetic field strength 
should reduce by such a factor during the phase of active accretion.  
We propose an extremely simple and elegant explanation for this
effect. It is seen that the screening becomes much weaker after 
the accretion has become isotropic. Hence, effectively, the magnetic 
field decreases until it is weak enough for the polar cap to open up 
completely to $90^0$ and then the field reduction slows down. It follows
from equation (\ref{eq_ct}) that the ratio of the initial to the final
field strength should be of the order of 
$\sim (\sin 90^0/ \sin \theta_{\rm min})^{7/2}$. Then, if 
$\theta_{\rm min}$ is about $5^0 - 10^0$, we obtain a factor of 
of $10^3$ -- $10^4$ between the initial and the final surface field.  
Therefore, if the accretion onto a neutron star starts with a polar
cap width of about $5^0 - 10^0$, the decrease in the magnetic field
comes out to be of the right order of magnitude.  
For the case presented in Figs.~\ref{fig05} and \ref{fig06}, we have 
taken $\theta_{\rm min} = 0.2$ such that the factor 
$\sin^{-7/2} \theta_{\rm min}$ is $\sim 2.83 \times 10^2$. 
Notice that in Fig.~\ref{fig05} the magnetic field decays approximately 
by the logarithm of this factor ($\sim 2.45$) before the decay essentially 
stops.

It should be noted here that the argument offered above does not 
account for the fact that the field decay continues via ohmic
dissipation, if the currents reside in the crustal region, even after
the accretion has become purely spherical \cite{kb97}. Detailed
modeling by \cite{kb99a} has shown that the ohmic dissipation is
effectively stalled and the surface field strength reaches an
asymptotic value only when material of mass $\Delta M \sim 10^{-2} \msun$ is
accreted. Therefore, if the time-scales of the opening of the polar 
cap is comparable to that of the accretion of $\Delta M$, the above 
argument would have a stronger claim to validity. However, a more detailed
calculation of the screening process, taking into account the
micro-physics of the neutron star crust, is required for a definite
answer in this connection. 

\section{Narrow Crustal Region {\label{scrust}}}

It has already been mentioned that all our previous calculations
are done by assuming the material flow happening in a wide subsurface
region, with the bottom boundary at $r_b = 0.5$, than what is 
expected in reality. This is done for the ease of numerical calculation.
However, in order to have an idea of the field evolution 
in such a narrow region here we reproduce some of the calculations of \S4
confining the initial field configuration as well as the material flow
in a region defined by $r_b = 0.90$ and $r_m = 0.95$, starting with
an initial configuration shown in Fig.~\ref{fig07}. But, it should be 
noted that in a real neutron star the flow region is confined to 
$r_m = 0.99$ which is beyond the viability limit of our numerical code. 

Fig.~\ref{fig08} shows the evolution of the magnetic field at a 
point on the surface at the mid-latitude with time. It is clear that 
the decay is much faster compared to that in the situation where the
flow happens in a much deeper region (Fig.~\ref{fig03}). This is due
to the fact that the distortion of the field lines occurs in a narrower 
region allowing for faster dissipation of the field through ohmic 
dissipation as can be interpreted from Fig.~\ref{fig07}. Moreover, the 
field saturates at a value $\sim 2.5$ order of magnitude smaller than 
the original, which is about the same as what we see in Fig.~\ref{fig05}. 
This lends credence to our argument that the field reduction basically 
depends on a geometric factor arising out of the polar cap opening and 
is independent of the depth of flow region. By extrapolating the results 
of this and the previous section, we expect that even in the case of a 
very narrow flow region appropriate for a neutron star, the magnetic 
field would first decay rapidly by a factor similar to what we have 
found above and would then saturate to an asymptotic value.

\section{Velocity without Counterflow {\label{scount}}}

\bef
\begin{center}{\mbox{\epsfig{file=fig09.ps,width=200pt,angle=-90}}}\end{center}
\caption[]{Material velocity without a region of interior reverse flow.}
\label{fig09}
\eef

\ni It needs to be mentioned here that there is no clear consensus
regarding the nature of the material flow. Many are of the opinion 
that there should be no counterflow and the flow in the surface layer 
should smoothly go over to a radial flow in the inner region (Geppert
U., Konenkov D., Spruit H.~C., {\it private communication}). The 
analytical form of the flow velocity, proposed by us, encompasses this 
possibility too. In the limit of $r_m \rightarrow r_b$ the flow takes 
this particular form as can be seen from Fig.~\ref{fig09}. 

\bef
\begin{center}{\mbox{\epsfig{file=fig10.ps,width=150pt,angle=-90}}}\end{center}
\caption[] {Evolution of the mid-latitude surface field with time for
a time-independent flow pattern. The curves marked $a,b$ correspond
to a material movement with and without (Fig.~\ref{fig09}) a reverse
flow. $\eta=0.05$ and $v_{\rm mb} = 0$ has been assumed for both the
cases.}
\label{fig10}
\eef

\bef
\begin{center}{\mbox{\epsfig{file=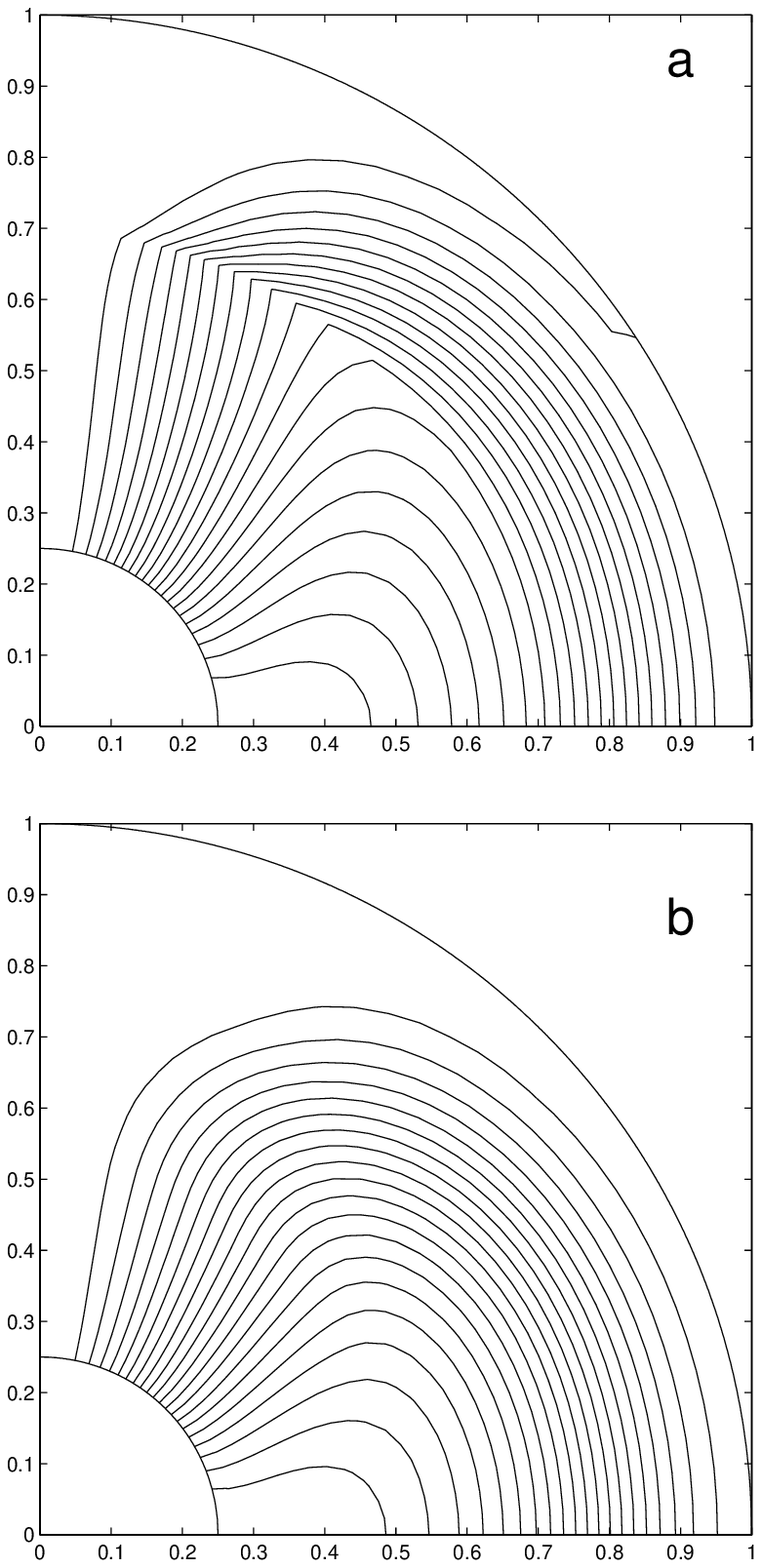,width=200pt}}}\end{center}
\caption[]{Internal field configuration at $t = 0.1$  starting from 
the initial configuration shown in Fig.~\ref{fig02}. The panel marked
$a$ corresponds to the velocity without a reverse flow shown in 
Fig.~\ref{fig09} whereas the configuration in panel $b$ has been evolved
with a velocity pattern with a reverse flow (panel $1a$ of Fig.\ref{fig01}).}
\label{fig11}
\eef

Fig.~\ref{fig10} shows the evolution of the magnetic field at a point 
on the surface at the mid-latitude with time. Comparing the case without 
a counterflow to the case with such a counterflow, it is evident that 
the decay is slower in the first case. Since, in absence of a counterflow 
in the region immediately beneath the surface layer, the stretching and 
the subsequent distortion in the magnetic field is less, the effective 
decrease in the surface field strength is also expected to be smaller 
than in the case with a counterflow present. Moreover, as can be seen 
in Fig.~\ref{fig11}, a flow pattern without an intermediate region of 
counterflow seem to give rise to a sharp bunching in the internal field 
configuration characteristic of the region in which the tangential 
surface flow changes over to a radial flow. This kind of bunching may 
perhaps lead to a faster dissipation of the currents at later times which
we have not investigated at present. To sum up, the evolution of the 
magnetic field is qualitatively similar even when there is no counterflow, 
although the factor by which the magnetic field decreases is somewhat smaller.

\section{Conclusions}

Many of the questions regarding the screening of the surface 
magnetic field of a neutron star still remain unanswered. We 
have tried to address this issue by presenting the first 
two-dimensional analysis of the problem. In our previous work, 
we have considered a steady flow pattern. In the present paper, 
we extend this to encompass the more realistic case of the accretion 
flow becoming more and more spherical with a reduction in the 
surface field strength. Therefore, with this work, the mathematical 
formulation of the problem can be said to be complete. However, 
a more detailed investigation of this problem incorporating the 
internal structure and the micro-physics of the neutron star crust 
is needed, and we expect to present the results of that investigation 
in a future communication \cite{sk03}. 

The main conclusion of this work is that the diamagnetic screening 
of the surface magnetic field becomes progressively less effective 
with time. As the strength of the surface field decreases due to 
screening, the magnetic field is less efficient in channeling
the material flow and the accretion becomes more and more radial. Such
radial in-fall of matter does not drag and bury the field lines
unlike in the case of a horizontal flow. This fact is borne out by our 
calculations, and it is seen that the decay effectively stops after the 
polar cap has fully opened up. We estimate that the screening by 
accreting material should reduce the field by a geometric factor of 
about $\sim 10^3 - 10^4$ reaching an asymptotic value thereafter. Happily, 
these numbers match with what is observed in binary and millisecond 
pulsars (compared to the field strength in normal pulsars), providing 
motivation for further investigations in this direction.

\section*{Acknowledgments}
We would like to thank U.~R.~M.~E. Geppert, Dennis Konenkov and
H. Spruit for suggesting the alternative form of the velocity field.
Both of us wish to thank IUCAA, Pune, where much of this work was 
carried out when SK was a post-doctoral fellow there and ARC was a
senior associate.

\bibliography{mnrasmnemonic,tap}
\bibliographystyle{mnras}

\bsp

\label{lastpage}

\end{document}